\documentclass[sn-basic,Numbered]{sn-jnl}% Math and Physical Sciences Reference Style
\usepackage{graphicx}%

\usepackage{amsmath,amssymb,amsfonts}%
\usepackage{amsthm}%
\usepackage{mathrsfs}%
\usepackage[title]{appendix}%
\usepackage{xcolor}%
\usepackage{textcomp}%
\usepackage{manyfoot}%
\usepackage{booktabs}%
\usepackage{algorithm}%
\usepackage{algorithmicx}%
\usepackage{algpseudocode}%
\usepackage{listings}%

\usepackage{multirow}
\usepackage{multicol}

\theoremstyle{thmstyleone}%
\theoremstyle{thmstyletwo}%

\theoremstyle{thmstylethree}%

\begin{document}

\title[Human in the AI loop via xAI and Active Learning for Visual Inspection]{Human in the AI loop via xAI and Active Learning for Visual Inspection}


\author*[1,5]{\fnm{Jo\v{z}e M.} \sur{Ro\v{z}anec}}\email{joze.rozanec@ijs.si}
\author[2]{\fnm{Elias} \sur{Montini}}
\author[2]{\fnm{Vincenzo} \sur{Cutrona}}
\author[3]{\fnm{Dimitrios} \sur{Papamartzivanos}}
\author[4]{\fnm{Timotej} \sur{Klemen\v{c}i\v{c}}}
\author[5]{\fnm{Bla\v{z}} \sur{Fortuna}}
\author[1]{\fnm{Dunja} \sur{Mladeni\'{c}}}
\author[6]{\fnm{Entso} \sur{Veliou}}
\author[3]{\fnm{Thanassis} \sur{Giannetsos}}
\author[7]{\fnm{Christos} \sur{Emmanouilidis}}

\affil*[1]{\orgname{Jo\v{z}ef Stefan Institute}, \orgaddress{\country{Slovenia}}}
\affil[2]{\orgname{University of Applied Sciences and Arts of Southern Switzerland}, \country{Switzerland}}
\affil[3]{\orgname{Ubitech Ltd.}, \orgaddress{\country{Greece}}}
\affil[4]{\orgname{University of Ljubljana}, \orgaddress{\country{Slovenia}}}
\affil[5]{\orgname{Qlector d.o.o.}, \orgaddress{\country{Slovenia}}}
\affil[6]{\orgname{University of West Attica}, \orgaddress{\country{Greece}}}
\affil[7]{\orgname{University of Groningen}, \orgaddress{\country{The Neatherlands}}}


\abstract{Industrial revolutions have historically disrupted manufacturing by introducing automation into production. Increasing automation reshapes the role of the human worker. Advances in robotics and artificial intelligence open new frontiers of human-machine collaboration. Such collaboration can be realized considering two sub-fields of artificial intelligence: active learning and explainable artificial intelligence. Active learning aims to devise strategies that help obtain data that allows machine learning algorithms to learn better. On the other hand, explainable artificial intelligence aims to make the machine learning models intelligible to the human person. The present work first describes Industry 5.0, human-machine collaboration, and state-of-the-art regarding quality inspection, emphasizing visual inspection. Then it outlines how human-machine collaboration could be realized and enhanced in visual inspection. Finally, some of the results obtained in the EU H2020 STAR project regarding visual inspection are shared, considering artificial intelligence, human digital twins, and cybersecurity.}


\maketitle

\section{Introduction}\label{S:INTRODUCTION}
% What is Industry 5.0 and how industry work profiles are changing?

Industrial revolutions have historically disrupted manufacturing by introducing automation into the production process.  
Increasing automation changed worker responsibilities and roles. While past manufacturing revolutions were driven from the optimization point of view, the Industry 5.0 concepts capitalize on the technological foundations of Industry 4.0 to steer manufacturing towards human-centricity \cite{EESC2018I5,longo2020value}, adding resilience and sustainability among its key targets \cite{EC2021I5}. This change is part of a holistic understanding of the industry's societal role. In particular, the European Commission expects the industry to collaborate on achieving societal goals that transcend jobs and company growth. 

Human-centric manufacturing within the Industry 5.0 aims to ensure that human well-being, needs, and values are placed at the center of the manufacturing process. Furthermore, it seeks to enable collaborative intelligence between humans and machines to enable co-innovation, co-design, and co-creation of products and services \cite{leng2022industry}, thus allowing leveraging on their strengths to maximize individual and joint outcomes and their joint added value \cite{Emmanouilidis2021Human}. It is expected that synergies enabled within Industry 5.0 will still allow for high-speed and mass-personalized manufacturing but will shift repetitive and monotonous tasks to be more assigned to machines to capitalize more on the human propensity for critical thinking and give them to more cognitively demanding tasks \cite{maddikunta2022industry}. 

The emerging shift in human roles goes beyond allowing them to move away from repetitive tasks to undertake other physical activities. As non-human actors, including artificial intelligence (AI) - enabled ones, undertake tasks that can be automated, humans are not necessarily excluded but may well play a higher added value and steering role, bringing their cognitive capabilities into the AI loop \cite{Emmanouilidis2019}. This includes active synergies between AI-enabled non-human entities and humans, resulting in novel work configurations \cite{Grønsund2020}. Such configurations empower human actors in new roles rather than diminishing them \cite{Amershi2014}. As a consequence, it is increasingly recognized that involving instead of replacing the human from the AI loop not only elevates the role of humans in such work environments but significantly enhances the machine learning process, and therefore the emergent capabilities of the AI-enabled actors \cite{Mosqueira2022}. As a result, such synergies involve humans and non-human entities who jointly contribute to shaping an emergent meta-human learning system, which in turn is more capable and powerful than human and non-human entities acting alone \cite{Lyytinen2020}.

A possible realization of such human-machine collaboration emerges from two sub-fields of artificial intelligence: active learning and explainable artificial intelligence (XAI). Active learning is concerned with finding pieces of data that allow machine learning algorithms to learn better toward a specific goal. Human intervention is frequently required, e.g., to label selected pieces of data and enable such learning. On the other hand, XAI aims to make the machine learning models intelligible to the human person so that humans can understand the rationale behind machine learning model predictions. While active learning requires human expertise to teach machines to learn better, XAI aims to help humans learn better about how machines learn and think. This way, both paradigms play on the strengths of humans and machines to realize synergistic relationships between them.

Among the contributions of the present work are (i) a brief introduction to the state-of-the-art research on human-machine collaboration, key aspects of trustworthiness and accountability in the context of Industry 5.0, and research related to automated visual inspection; (ii) the development of a vision on how an AI-first human-centric visual inspection solution could be realized; and (iii) a description of experiments and results obtained in the field of automated visual inspection at the EU H2020 STAR project.

% How can it be realized? Paradigms and applications.
% Introduce examples of such humans in the AI loop involvement and zoom in to Active Learning

% Why is it relevant? Provide some perspective from multiple points of view.
% Why is it relevant for Manufacturing? Zoom in to Quality Control. Then it would be a natural narrative to introduce the paper's contribution.

% Open challenges? Shall we leave that for the discussion?

% Chapter structure
The rest of the work is structured as follows: Section~\ref{S:BACKGROUND} describes related work, providing an overview of human-machine collaboration, the industry 5.0 paradigm and human-centric manufacturing, state-of-the-art on automated quality inspection, and a vision of how human-machine collaboration can be realized in the visual inspection domain. In Section \ref{S:USE-CASES}, relevant research contributions from the EU H2020 STAR project are outlined, offering concrete examples of humans and AI working in synergy. Finally, Section~\ref{S:CONCLUSION} provides conclusions and insight into future work.

\section{Background}\label{S:BACKGROUND}

\subsection{Overview on Human-Machine Collaboration}\label{SS:HUMAN-MACHINE-COLLABORATION}

% What is a machine? What kind of interactions do we know?
The advent of increasingly intelligent machines has enabled a new kind of relationship: the relationship between humans and machines. Cooperative relationships between humans and machines were envisioned back in 1960 \cite{licklider1960man,gerber2020conceptualization}. This work defines machines in a broad sense, considering intelligent systems that can make decisions autonomously and independently (e.g., automated, autonomous, or AI agents, robots, vehicles, and instruments) \cite{rahwan2022machine,xiong2022challenges,gerber2020conceptualization}. Relationships between humans and machines have been characterized through different theories, such as the Socio-Technical Systems theory (considers humans and technology shape each other while pursuing a common goal within an organization), Actor-Network Theory (considers machines should be equally pondered by humans when analyzing a social system, considering the later as an association of heterogeneous elements), Cyber-Physical Social Systems theory (extends the Socio-Technical Systems theory emphasizing social dimensions where computational algorithms are used to monitor devices), the theory on social machines (considers systems that combine social participation with machine-based computation), and the Human-Machine Networks theory (considers humans and machines form interdependent networks characterized by synergistic interactions). The first three theories conceptualize humans and machines as a single unit, while the last two consider social structures mediated in human-machine networks. In particular, the Socio-Technical Systems theory considers humans and technology shape each other while pursuing a common goal within an organization. The Cyber-Physical Social Systems theory extends this vision, emphasizing social dimensions where computational algorithms are used to monitor and control devices. Moreover, the Actor-Network Theory conceptualizes the social system as an association of heterogeneous elements and advocates that machines should be equally pondered to humans. The theory of social machines is interested in systems that combine social participation with machine-based computation. In contrast, the Human-Machine Networks theory considers humans and machines to form interdependent networks characterized by synergistic interactions. A thorough analysis of the abovementioned concepts can be found in \cite{tsvetkova2017understanding}.

% Symbiosis. What do we expect?
Regardless of the particular theory, the goal remains the same: foster and understand mutualistic and synergistic relationships between humans and machines, where the strengths of both are optimized towards a common goal to achieve what was previously unattainable to each of them. To that end, individual roles must either be clearly defined or allow for a clear sliding of roles when a role can be shared among different types of actors. This will ensure a dynamic division of tasks, optimal use of resources, and reduced processing time. Machines are aimed at supporting, improving, and extending human capabilities. The joint outcomes of human-machine collaboration can result in systems capable of creativity and intuitive action to transcend mere automation. Communication is a critical aspect of every social system. Therefore, emphasis must be placed on the interaction interfaces between such actors. To make such interfaces effective, the concept of shared context or situation awareness between collaborating agents becomes essential and can be seen as a form of mutual understanding \cite{Emmanouilidis2019}. This shared context is enabled through interaction communication of different modalities, including direct verbal (speech, text) and non-verbal (gestures, action and intention recognition, emotions recognition). On the other hand, means must be designed so that humans can understand the machine's goals and rationale for acting to reach such goals in a human-like form. In this regard, human-machine interfaces to support multi-modal interaction play a crucial role. These aspects were also identified by Jwo et al. \cite{jwo2021smart}, who described the 3I (Intellect, Interaction, and Interface) aspects that must be considered for achieving human-in-the-loop smart manufacturing.

% How can we enhance symbiosis/complementarity?
Beyond shared context, human-machine cooperation requires adequate communication and shared or sliding control \cite{Tang2016sliding}. To realize an effective bidirectional information exchange, theory and methods must address how data and machine reasoning can be presented intuitively to humans. Frameworks and models abstracting human cognitive capabilities \cite{Langley2017cognitive}
are key to achieving this. Aligning the design of interactive interfaces and support tools for human-machine interactions with such concepts can be critically important for making effective human–machine interfaces. Enhancements in the interactivity, multisensitivity, and autonomy of feedback functions implemented on such interfaces allow for deeper integration between humans and machines. Shared control can be articulated at operational, tactical, and strategic levels, affecting information-gathering, information-analysis, decision-making, and action implementation.

% What kind of roles do they uptake? What kind of problems do arise? Responsbility, bias, etc.
Human-machine interactions can be viewed from multiple perspectives, necessitating a thorough consideration of several factors influencing such collaborations. These factors encompass emotional and social responses, task design and assignment, trust, acceptance, decision-making, and accountability \cite{chugunova2020we}. 
Notably, research indicates that machines in collaborative settings impact human behavior, resulting in a diminished emotional response toward them. Consequently, this reduced emotional response can foster more rational interactions. Moreover, studies reveal that humans perceive a team more favorably when machines acknowledge and admit their errors. Additionally, the absence of social pressure from humans can detrimentally affect overall human productivity. Furthermore, concerning accountability for decision-making, humans tend to shift responsibility onto machines.

Trust, a critical aspect to consider, has been explored extensively. Studies demonstrate that trust in machines is closely linked to perceived aptness \cite{chugunova2020we}. Instances of machine errors often lead to a loss of trust, particularly when machines act autonomously. However, if machines operate in an advisory capacity, trust can be amended over time. Additionally, research reveals that while humans value machine advice, they hesitate to relinquish decision-making authority entirely. Nevertheless, relying excessively on machines can result in sub-optimal outcomes, as humans may fail to identify specific scenarios that necessitate their attention and judgment. For further details about the abovementioned experiments and additional insights the reader may be interested on the works by Chugunova et al. \cite{chugunova2020we}.

\subsection{Industry 5.0 and Human-Centric Manufacturing}\label{SSS:INDUSTRY50}

\subsubsection{New Technological Opportunities to Reshape the Human Workforce}\label{SSS:INDUSTRY50-RESHAPING-WORKFORCE}

Digital transformation in production environments demands new digital skills and radically reshapes the roles of plant and machine operators \cite{tschang2021artificial,chuang2022indispensable}. While Industry 4.0 emphasizes the use of technologies to interconnect different stages of the value chain and the use of data analytics to increase productivity, Industry 5.0 emphasizes the role of humans in the manufacturing context \cite{EUindustry5,kaasinen2022smooth}. Furthermore, it aims to develop means that enable humans to work alongside advanced technologies to enhance industry-related processes \cite{lu2021current}. An extensive review of this paradigm and its components was written by Leng et al. \cite{leng2022industry}. Nevertheless, two components are relevant to this work: Collaborative Intelligence and Multi-objective Interweaving. Collaborative Intelligence is the fusion of human and AI \cite{wilson2018collaborative}. In the context of Industry 5.0, the fusion of both types of intelligence entails the cognitive coordination between humans and AI in machines, enabling them to collaborate in the innovation, design, and creation of tailored products and services. Complementarity between humans and AI (see Table \ref{T:HUMANS-VS-AI}) leads to the more effective execution of such tasks than would be possible if relegated to humans or machines only \cite{montini2023,cai2019human,rovzanec2023predicting,jarrahi2022key}. 

When analyzing complementarities, humans have the knowledge and skills to develop and train machines by framing the problems to be solved and providing feedback regarding their actions or outputs \cite{jarrahi2018artificial,paul2022intelligence,leng2022industry,WU202375}. Furthermore, humans can enrich machine outcomes by interpreting results and insights and deciding how to act upon them \cite{WU2022364}. Machines amplify workers' cognitive abilities: they can track many data sources and decide what information is potentially relevant to humans. Furthermore, machines can excel at repetitive tasks and free humans from such a burden. Such complementary is considered within the multi-objective interweaving nature of Industry 5.0, which enables optimizing multiple goals beyond process performance and social and environmental sustainability \cite{bettoni2020mutualistic}. Moreover, research suggests that leading companies are beginning to recognize the benefit of using machines and automation systems to supplement human labor rather than replacing the human workforce entirely \cite{Accenture2018,Deloitte2018}.
While AI was already able to tackle certain tasks with super-human capability \cite{ciregan2012multi}, it has recently shown progress in areas such as creativity (e.g., through generative models such as DALL·E 2 \cite{ramesh2022hierarchical}) or problem-solving \cite{cao2022new}, opening new frontiers of human-machine collaboration, such as co-creativity \cite{liapis2016can,anantrasirichai2022artificial}.

In addition to the direct human involvement described above, digital twins \cite{montini2022iiot} are another way to incorporate human insights into the AI processes. By creating virtual models of human behaviour and mental processes, more profound insights into how humans interact with the world and use this information to improve AI systems. Digital twins can also support explainability and transparency in AI systems, making explaining how they arrive at their decisions easier \cite{bansal2019beyond}. Moreover, digital representations can be used to consider users' preferences in the AI system behaviours, e.g., type of support \cite{van2021human,hu2022review}.

\begin{table}[htb!]
\centering
\begin{tabular}{|ll|c|c|}
\hline
\multicolumn{2}{|c|}{\textbf{Capability}} & \multicolumn{1}{l|}{\textbf{Humans}} & \multicolumn{1}{l|}{\textbf{Machines}} \\ \hline
\multicolumn{1}{|l|}{\multirow{17}{*}{\textbf{Strengths   and capabilities}}} & leadership & x &  \\ 
\multicolumn{1}{|l|}{} & teamwork & x &  \\ 
\multicolumn{1}{|l|}{} & creativity & x & o \\ 
\multicolumn{1}{|l|}{} & problem-solving & x & x \\ 
\multicolumn{1}{|l|}{} & risk assessment & x & o \\ 
\multicolumn{1}{|l|}{} & intuition & x &  \\ 
\multicolumn{1}{|l|}{} & interpretation & x &  \\ 
\multicolumn{1}{|l|}{} & empathy & x &  \\ 
\multicolumn{1}{|l|}{} & adapt behaviour & x & o \\ 
\multicolumn{1}{|l|}{} & learn from experience & x & o \\ 
\multicolumn{1}{|l|}{} & speed &  & x \\ 
\multicolumn{1}{|l|}{} & scalability &  & x \\ 
\multicolumn{1}{|l|}{} & endurance &  & x \\ 
\multicolumn{1}{|l|}{} & quantitative accuracy &  & x \\ 
\multicolumn{1}{|l|}{} & process large amounts of data &  & x \\  
\multicolumn{1}{|l|}{} & process different kinds of data in parallel & x &  \\ 
\multicolumn{1}{|l|}{} & perform continuous operations &  & x \\  
\multicolumn{1}{|l|}{} & consistent decision-making &  & x  \\ 
\multicolumn{1}{|l|}{} & physical and cognitive abilities & x & o\\ \hline
\multicolumn{1}{|l|}{\multirow{9}{*}{\textbf{Weaknesses}}}   
& prone to biases and errors & x &  \\ 
\multicolumn{1}{|l|}{} & affected by emotions & x &  \\ 
\multicolumn{1}{|l|}{} & affected by distractions & x &  \\ 
\multicolumn{1}{|l|}{} & prone to frauds and adversarial attacks & o & x \\ 
\multicolumn{1}{|l|}{} & affected by fatigue & x &  \\ 
\multicolumn{1}{|l|}{} & limited to certain scope and goals &  & x \\ 
\multicolumn{1}{|l|}{} & lack of emotional intelligence & & x \\  
\multicolumn{1}{|l|}{} & lack of social skills &  & x \\ \hline
\end{tabular}
\caption{Overview on humans and AI complementarity (adapted from
%Jarrahi et al., Paul et al., and Leng et al.
\cite{jarrahi2018artificial,paul2022intelligence,leng2022industry}, and complemented with our observations). \textbf{x}: capability completely fulfilled; \textbf{o}: capability partially fulfilled.}
\label{T:HUMANS-VS-AI}
\end{table}

\subsubsection{Trustworthiness and Implications for AI-driven Industrial Systems}\label{SSS:TRUSTWORTHINESS-AND-AI}

Trustworthiness for systems and their associated services and characteristics is defined according to the International Organization for Standardization (ISO) as ``\emph{the ability to meet stakeholders' expectations in a verifiable way}'' \cite{ISO2022Trustworthiness}. It follows that trustworthiness can refer to products, services, technology, and data and, ultimately, to organizations. Therefore, the concept of trustworthiness is directly applicable to AI-driven systems, particularly to human-centric AI-enabled solutions. However, it should be understood that trustworthiness is a multifaceted concept, incorporating distinct characteristics such as accountability, accuracy, authenticity, availability, controllability, integrity, privacy, quality, reliability, resilience, robustness, safety, security, transparency, usability \cite{ISO2022Trustworthiness}.

Some of these characteristics should be seen as emerging characteristics of AI-enabled systems, which are not solely determined by the AI's contribution to an overall solution. Focusing specifically on the AI components of such solutions, ethics guidelines published by the European Commission (EC) identifies seven key requirements for trustworthiness characteristics that must be addressed \cite{EUTrustworthy}. These include (i) human agency and oversight, (ii) technical robustness and safety, (iii) privacy and data governance, (iv) transparency, (v) diversity, non-discrimination, and fairness, (vi) societal and environmental well-being, and (vii) accountability. Regarding some of these characteristics, there is a direct correspondence between broader trustworthiness as documented according to ISO and the EC guidelines. Technical robustness, safety, privacy, transparency, and accountability are identified in both sources. Human agency and oversight are directly linked to controllability, and so is governance, which is also the prime focus of ISO recommendations \cite{ISO2015governance}. Given the societal impacts that AI-induced outcomes can have, the EC has also highlighted diversity, non-discrimination, fairness, and societal and environmental well-being as key characteristics of trusted AI solutions. However, these aspects are also partly addressed as part of the broader concept of "freedom from risk", which can be defined as the extent to which a system avoids or mitigates risks to economic status, human life, health, and well-being and or the environment \cite{ISO2016FreedomFromRisk}.

The trustworthiness of an AI system can be affected by multiple factors. Some of them relate to cybersecurity. In particular, machine learning algorithms are vulnerable to poison and evasion attacks. During poisoning attacks, the adversary aims to tamper with the training data used to create the machine learning models and distort the AI model on its foundation \cite{gu2017badnets,shokri2020bypassing}. Evasion attacks are performed during inference, where the attacker crafts adversarial inputs that may seem normal to humans but drive the models to classify the inputs wrongly \cite{madry2017towards,10.1145/3134600.3134635}. Such an adversarial landscape poses significant challenges and requires a collaborative approach between humans and machines to build defenses that can lead to more robust and trustworthy AI solutions. While human intelligence can be used for the human-in-the-loop adversarial generation, where humans are guided to break models \cite{wallace2019trick}, AI solutions can be trained to detect adversarial inputs and uncover potentially malicious instances that try to evade the AI models \cite{s22186905}. Furthermore, human-machine collaboration can be fostered to detect such attacks promptly.

Accountability refers to the state of being accountable and relates to allocated responsibility \cite{ISO2022Trustworthiness}. At the system level, accountability is a property that ensures that the actions of an entity can be traced uniquely to the entity \cite{ISO1989Security}. However, when considering governance, accountability is the obligation of an individual or organization to account for its activities, accept responsibility for them, and disclose the results in a transparent manner \cite{ISO2015governance}. Therefore accountability is closely linked to transparency for AI-enabled systems, which is served via XAI and interpretable AI. XAI and interpretable AI ensure that AI systems can be trusted when analyzing model outcomes that impact costs and investments or whenever their outputs provide information to guide human decision-making. Accuracy generally refers to the closeness of results and estimates to true values but, in the context of AI, further attains the meaning appropriate for specific machine learning tasks. Any entity that is what it claims to be is said to be characterized by authenticity, with relevant connotations for what AI-enabled systems claim to deliver. Such systems may furthermore be characterized by enhanced availability to the extent that they are usable on demand. Other characteristics such as integrity, privacy, and security attain additional meaning and importance in AI-driven systems and are further discussed in the next section. They can contribute to and affect the overall quality, reliability, resilience, robustness, and safety, whether the unit of interest is a component, a product, a production asset, or a service, with implications for individual workers all the way to the organization as a whole.
When considering accountability for AI systems from the legal perspective, the EU AI Act \cite{AIact} in its current form considers developers and manufacturers responsible for AI failures or unexpected outcomes. Nevertheless, the concept of accountability will evolve based on the issues found in practice and the corresponding jurisprudence that will shape the learning on how different risks, contexts, and outcomes must be considered in the industry context \cite{hohma2023investigating}.

\subsection{Automated Quality Inspection}\label{SS:QUALITY-INSPECTION-OVERVIEW}

\subsubsection{The Role of Robotics}\label{SSS:ROBOTICS}

The increasing prevalence of human-robot collaboration in diverse industries showcases the efforts to enhance workplace productivity, efficiency, and safety through the symbiotic interaction of robots and humans \cite{heyer2010human}. In manufacturing, robots are employed for repetitive and physically demanding tasks, enabling human workers to allocate their skills toward more intricate and creative endeavors. This collaborative partnership allows for the fusion of human and robot capabilities, maximizing the overall outcomes.

The successful implementation of human-robot interaction owes credit to collaborative robots, commonly called cobots \cite{kosuge2004human}. These advanced robots have sophisticated sensors and programming that facilitate safe and intuitive human interaction. This collaboration improves productivity and fosters a work environment where humans and robots can coexist harmoniously. This approach harmoniously merges robots' precision and accuracy with human workers' adaptability and dexterity.

Robotic integration in product quality control has become widespread across diverse industries and production sectors. Robots offer exceptional advantages within quality inspection processes, including precise repeatability and accurate movements \cite{brito2020machine}. They possess the capability to analyze various product aspects such as dimensions, surface defects, color, texture, and alignment, ensuring adherence to predefined standards. Robots' superior accuracy and efficiency make them an ideal choice for quality control applications.

To facilitate quality testing, robots are equipped with a range of sensors. These sensors enable precise measurement, detection, and sorting operations. Robots with cameras utilize advanced machine vision techniques to analyze image and video streams and identify anomalies like cracks, scratches, and other imperfections \cite{villalba2019deep}. Subsequently, defective items are segregated from conforming ones, elevating overall production quality. The industry is witnessing an increasing adoption of 3D vision systems, particularly in applications requiring object grasping and precise information about object position and orientation.

Specially designed robots, such as coordinate measuring machines, are employed for dimensional and precision measurements. These robots feature high-precision axis encoders and accurate touch probes, enabling them to detect part measurements and consistently evaluate adherence to quality standards \cite{leach2019geometrical}.

The active learning paradigm can be applied to enable efficient and flexible learning in robots. This can be particularly useful in resource-constrained industrial environments, where data scarcity and limited human knowledge prevail, acquiring essential data through unsupervised discovery becomes imperative \cite{daniel2014active}. Active learning demonstrates extensive applicability in robotics, encompassing prioritized decision-making, inspection, object recognition, and classification. Within quality control, active learning algorithms optimize machine learning models' defect detection and quality assessment training process. By actively selecting informative samples for labeling, active learning minimizes labeling efforts, augments model training efficiency, and ultimately enhances the accuracy and performance of quality control systems.

An intriguing domain of investigation pertains to the advancement of intuitive and natural interfaces that foster seamless communication and interaction between humans and robots. This entails the exploration of innovative interaction modalities, encompassing speech, gestures, and facial expressions, or even using augmented reality to customize the robots' appearance and foster better interaction with humans \cite{lambert2020systematic}. Other key research areas involve developing adaptive and flexible robotic systems that dynamically adapt their behavior and actions to the prevailing context and the human collaborator's preferences, achieving low processing times \cite{mukherjee2022survey}. These could be critical to enable real-time human intent recognition, situational awareness, and decision-making, all aimed at augmenting the adaptability and responsiveness of robots during collaborative tasks.

\subsubsection{Artificial Intelligence - Enabled Visual Inspection}\label{SSS:QUALITY-INSPECTION-OVERVIEW}

Visual inspection is frequently used to assess whether the manufactured product complies with quality standards and allows for the detection of functional and cosmetic defects \cite{chin1982automated}. It has historically involved human inspectors in determining whether the manufactured pieces are defective. Nevertheless, the human visual system excels in a world of variety and change, while the visual inspection process requires repeatedly observing the same product type. Furthermore, human visual inspection suffers from poor scalability and the fact that it is subjective, creating an inherent inspector-to-inspector inconsistency. The quality of visual inspection can be affected by many factors. See \cite{see2012visual} classified them into five categories, whether they are related to the (i) task, (ii) individual, (iii) environment, (iv) organization, or (v) social aspects.

To solve the issues described above, much effort has been invested in automated visual inspection by creating software capable of inspecting manufactured products and determining whether they are defective. Cameras are used to provide visual input. Different approaches have been developed to determine whether a defect exists or not. 

Automated optical quality control may target visual features as simple as colors, but more complex ones are involved in crack detection, the orientation of threads, defects in bolts \cite{Rajan2020CNN} and metallic nuts \cite{Bharti2022Nuts}. Through automated optical inspection, it is also possible to detect defects on product surfaces of wide-ranging sizes \cite{Yun2020metal,Tsai2021Autoencoder,Cao2018defect}. Furthermore, it is also possible to target the actual manufacturing process, for example, welding \cite{Tripicchio2020Defect},  injection molding \cite{Liu2021moulding}, or assembly of manufactured components \cite{Frustaci2022}. Additionally, automated visual inspection applies to remanufacturing products at the end of their useful life \cite{Saiz2021remanufacturing}.

State-of-the-art (SOTA) automated visual inspection techniques are dominated by deep learning approaches, achieving high-performance levels \cite{aggour2019artificial}. Among the many types o learning from data for visual inspection, unsupervised, weakly supervised, and supervised methods can be named. Unsupervised methods aim to discriminate defective manufactured pieces without labeled data. The weakly-supervised approach assumes that data has an inherent cluster structure (instances of the same class are close to each other) and that the data lies in a manifold (nearby data instances have similar predictions). Therefore, it leverages a small amount of annotated data and unlabeled data to learn and issue predictions. Finally, supervised methods require annotated data and usually perform best among the three approaches. Often, labeled data are unavailable in sufficient range and numbers to enable fully supervised learning and additional exemplar images can be produced through data augmentation \cite{Kim2021overview}. In addition, multiple strategies have been developed to reduce the labeled data required to train and enhance a given classifier. Among them are active learning, generative AI, and few-shot learning. In the context of visual inspection, active learning studies how to select data instances that can be presented to a human annotator to maximize the models' learning. Generative AI aims to learn how to create data instances that resemble a particular class. Finally, few-shot learning aims to develop means by which the learner can acquire experience to solve a specific task with only a few labeled examples. To compensate for the lack of labeled data, it can either augment the dataset with samples from other datasets or use unlabeled data, acquire knowledge on another dataset, or algorithm (e.g., by adapting hyperparameters based on prior meta-learned knowledge) \cite{wang2020generalizing}.

Regardless of the progress made in automated visual inspection, many challenges remain. First, there is no universal solution for automated visual inspection: solutions and approaches have been developed to target a specific product. Flexibility to address the inspection of multiple manufactured products with a single visual inspection system is a complex challenge and remains an open issue \cite{chin1982automated,newman1995survey,czimmermann2020visual}. Second, unsupervised machine learning models do not require annotating data and may provide a certain level of defect detection when associating data clusters to categories (e.g., types of defects or no defects). Furthermore, given that no prior annotation of expected defects is required, they are suitable when various defects exist. Nevertheless, their detection rates are lower than those obtained by supervised machine learning models. Therefore, it should be examined use case by use case whether the unsupervised machine learning models are a suitable solution. Third, data collection and annotation are expensive. While data collection affects unsupervised machine learning models, data collection and annotations directly impact supervised machine learning approaches. While multiple strategies have been envisioned to overcome this issue (e.g., generative models, active learning, and few-shot learning), data collection and annotation remain an open challenge. Finally, better explainability techniques and intuitive ways to convey information to humans must be developed to understand whether the models learn and predict properly.

\subsection{Realizing Human-Machine Collaboration in Visual Inspection}\label{SS:QUALITY-INSPECTION-AI}

% Active Learning: short intro, why it is relevant, how it helps in this regard
% Abductive learning: Z.-H. Zhou. Abductive learning: Towards bridging machine learning and logical reasoning. SCIENCE CHINA Information Sciences, 62(7):76101:1–76101:3, 2019.
% XAI: short intro, why it is relevant, how it helps in this regard. XAI and metalearning.
% Few shot learning: short intro, why it is relevant, how it helps in this regard
% Do we have synergies between Active Learning, XAI, and few-shot learning?

While much progress has been made in automated visual inspection, authors recognize that most solutions are custom and developed for a particular product type. Developing systems that could adapt to a broad set of products and requirements remains a challenge. In human-centered manufacturing, it is critical to rethink and redesign the role of humans in the visual inspection process. The role of humans in automated visual inspection is shifting away from repetitive and manual tasks to roles with more cognitive involvement, which can still not be replicated by machines and AI. In the simplest case, this involves humans labeling acquired image samples to guide the machine learning process \cite{Silva2023Injection}. However, the role of humans extends beyond data labeling and may involve interaction loops between humans and AI as part of the machine learning process \cite{Marz2022Interactive}.

In this regard, two machine-learning paradigms are particularly important: active learning and XAI. On one side, active learning is an AI paradigm that seeks the  intervention of an oracle (usually a human person) to help the machine learning model learn better toward an objective. XAI, on the other side, aims to explain the rationale behind a machine learning model action or prediction. Doing so enables a fruitful dialogue between humans and machines by providing insights into the machines' rationale and decision-making process.

Active learning for classification is based on the premises that unlabeled data (either collected or generated) is abundant, the data labeling is expensive, and the models' generalization error can be minimized by carefully selecting new input instances with which the model is trained \cite{settles2009active,sugiyama2012active}. Active learning for classification has traditionally focused on the data (selecting or generating the data without further consideration for the model at hand) and the model's learning (e.g., considering the uncertainty at the predicted scores). Nevertheless, approaches have been developed to consider both dimensions and provide a holistic solution. One of them is the Robust Zero-Sum Game (RZSG) framework \cite{zhu2019robust}, which attempts to optimize both objectives at once, framing the data selection as a robust optimization problem to find the best weights for unlabeled data to minimize the actual risk, reduce the average loss (to achieve greater robustness to outliers) and minimize the maximal loss (increasing the robustness to imbalanced data distributions). Another perspective has been considered by Zajec et al. \cite{zajecetal} and Kri\v{z}nar et al. \cite{kriznaretal}, who aim to select data based on insights provided by XAI methods and therefore benefit from direct insights into the model's learning dynamics. Regardless of the approach, Wu et al. \cite{wu2018pool} propose that three aspects must be considered when searching for the most valuable samples: informativeness (contains rich information that would benefit the objective function), representativeness (how many other samples are similar to it), and diversity (the samples do not concentrate in a particular region, but rather are scattered across the whole space). Strategies will be conditioned by particular requirements (e.g., whether the data instances are drawn from a pool of samples or a data stream). For a detailed review of active learning, the reader may be interested in some high-quality surveys of this domain. In particular, the works by Settles \cite{settles2009active} and Ro\v{z}anec et al. \cite{futureDM2022} can serve as an introduction to this topic. Furthermore, the surveys by Fu et al. \cite{fu2013survey} and Kumar et al. \cite{kumar2020active} provide an overview of querying strategies in a batch setting; the survey by Lughofer \cite{lughofer2017line} give an overview of active learning in online settings, and the study by Ren et al. \cite{ren2021survey} describes active learning approaches related to deep learning models.

While AI models have the potential to automate many tasks and achieve super-human performance levels, in most cases, such models are opaque to humans: their predictions are mostly accurate, but no intuition regarding their reasoning process is conveyed to humans. Understanding the rationale behind a model's prediction is of utmost importance, given it provides a means to assess whether the predictions are based on accurate facts and intuitions. Furthermore, it is crucial to develop means to understand the model's reasoning process given the impact such techniques have on the real world, either in fully automated settings or when decision-making is delegated to humans. Such insights enable responsible decision-making and accountability. The subfield of AI research developing techniques and mechanisms to elucidate the models' rationale and how to present them to humans is known as XAI. While the field can be traced back to the 1970s \cite{scott1977explanation}, it has recently flourished with the advent of modern deep learning \cite{xu2019explainable}. When dealing with XAI, it is important to understand what makes a good explanation.  A good explanation must consider at least three elements \cite{arrieta2020explainable}: (a) reasons for a given model output (e.g., features and their values, how strongly do features influence a forecast, whether the features at which the model looks at make sense w.r.t. the forecast, how did training data influence the model's learning), (b) context (e.g., the data on which the machine learning model was trained, the context on which inference is performed), and (c) how is the abovementioned information conveyed to the users (e.g., target audience, the terminology used by such an audience, what information can be disclosed to it). XAI can be valuable in enhancing human understanding with new (machine-based) perspectives. It can also help to understand whether the model is optimizing for one or few of all required goals and therefore identify an appropriate compromise between the different goals that must be satisfied for the problem at hand \cite{doshi2017towards}. To assess the goodness of an explanation, aspects such as user satisfaction, the explanation persuasiveness, the improvement of human judgment, the improvement of human-AI system performance, the automation capability, and the novelty of explanation must be considered \cite{schwalbe2023comprehensive}.
For a detailed review of XAI, the reader may consider the works of Arrieta et al. \cite{arrieta2020explainable}, Doshi-Velez et al. \cite{doshi2017towards}, and Schwalbe et al. \cite{schwalbe2023comprehensive}. The work of Bodria et al. \cite{bodria2023benchmarking} provides a comprehensive introduction to XAI black box methods, and the works of Doshi-Velez et al. \cite{doshi2017towards}, Hoffman et al. \cite{hoffman2018metrics} and Das et al. \cite{das2020opportunities} focus on insights about how to measure the quality of explanations.

\begin{figure}
\begin{center}
\includegraphics[width=\textwidth]{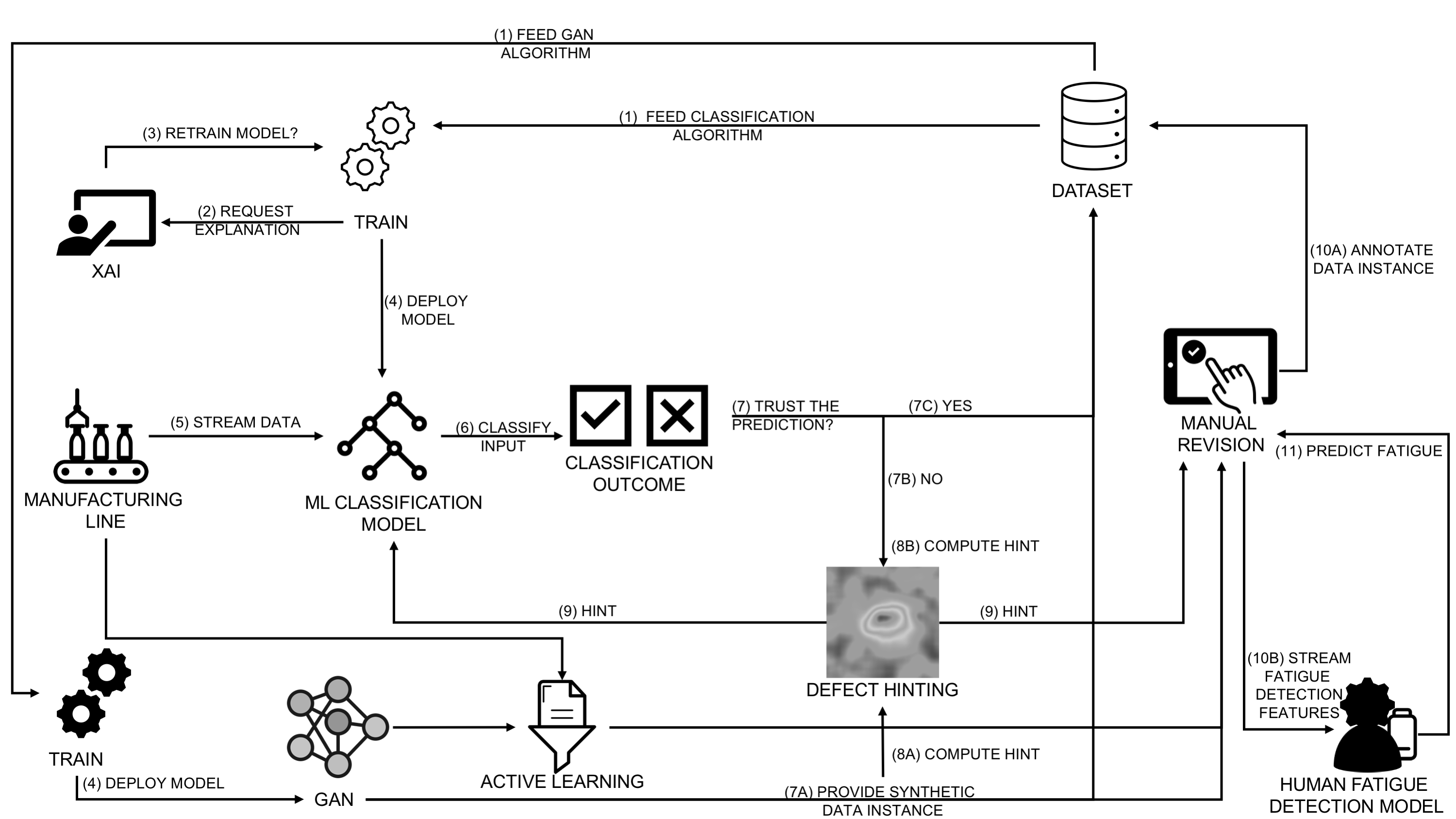}
\caption{Envisioned setup for an AI-first human-centric visual inspection solution.} 
\label{F:AI-FIRST-HUMAN-CENTRIC-VISUAL-INSPECTION}
\end{center}
\end{figure}

Active learning and XAI can complement each other. Understanding the rationale behind a model prediction provides valuable insight to humans and can also be leveraged in an active learning setting. In the particular case of defect inspection, insights obtained by XAI techniques are usually presented in anomaly maps. Such anomaly maps highlight regions of the image the machine learning models consider to issue a prediction. The more perfect the learning of a machine learning model, the better those anomaly maps should annotate a given image indicating defective regions. Therefore, the insights obtained from those anomaly maps can be used in at least two ways. First, the anomaly maps can be handed to the oracle (human inspector), who, aided by the anomaly map and the image of the product, may realize better where the manufacturing errors are, if any. Second, anomaly maps can be used to develop novel models and active learning policies that allow for data selection, considering what was learned by the model and how the model perceives unlabeled data. This approach is detailed in Fig. \ref{F:AI-FIRST-HUMAN-CENTRIC-VISUAL-INSPECTION}, which depicts how an initial dataset is used to train machine learning models for defect classification or data generation. In the model training process, XAI can be used to debug and iterate the model until getting satisfactory results. The classification model is then deployed to perform inference on incoming product images from the manufacturing line. If the classification scores for certain classes are high enough, the product can be classified as good or defective. When the uncertainty around the predicted scores is not low enough, the case can be sent for manual revision. Insights obtained through XAI and unsupervised classification models can be used to hint to the human inspector where the defects may be located. Alternative data sources for the manual revision or data labeling process can be generative models (e.g., generative adversarial networks), which can be used to generate labeled synthetic data and validate the level of attention of a human inspector. When collecting data, active learning techniques can be used to select the most promising data instances from either generative models or incoming images from the manufacturing line, reducing the labeling effort. Finally, a separate model can monitor human inspectors to predict fatigue and performance. Such models can be a valuable tool to ensure workplace well-being and enhance work quality. Some of the results obtained within the STAR project are presented in Section \ref{SS:VISUAL-INSPECTION}.

In recent years, researchers have made significant progress in understanding and quantifying fatigue and recognizing its impact on human performance and overall well-being. Through AI techniques, new approaches have emerged to accurately estimate the fatigue levels of individuals during different tasks and in different contexts \cite{hooda2022comprehensive,aguirre2021machine}.
One notable area of inquiry concerns the assessment of fatigue in the workplace. Understanding and managing worker fatigue has become essential given the increasing demands and pressures of modern work environments. AI models can consider various factors and features to assess employee fatigue levels accurately. These models can provide valuable insights for organizations looking to implement strategies and interventions to optimize productivity and ensure employee well-being or to support workflows, including quality controls, such as identifying when operators need a break. Although laboratory experiments have been conducted in this area \cite{leone2020multi}, industrial applications remain relatively restricted compared to other fields, such as driving \cite{sikander2018driver}.

\section{Industrial applications}\label{S:USE-CASES}

This section briefly describes how some ideas presented in the previous sections have been realized within the EU H2020 STAR project. Three domains are considered: artificial intelligence for visual inspection, digital twins, and cybersecurity.

\subsection{Machine Learning and Visual Inspection}\label{SS:VISUAL-INSPECTION}
In the domain of visual inspection, multiple use cases were considered. The datasets were provided by two industrial partners: \textit{Philips Consumer Lifestyle BV} (Drachten, The Netherlands) and \textit{Iber-Oleff - Componentes Tecnicos Em Pl\'astico, S.A.} (Portugal). The \textit{Philips Consumer Lifestyle BV} manufacturing plant is considered one of Europe's most important Philips development centers and is devoted to producing household appliances. They provided us with three datasets corresponding to different products. The first one corresponded to logo prints on shavers. The visual inspection task required understanding whether the logo was correctly printed or had some printing defect (e.g., double printing or interrupted printing). The second one corresponded to decorative caps covering the shaving head's center, and it required identifying whether the caps were correctly manufactured or if some flow lines or marks existed. Finally, the third dataset was about toothbrush shafts transferring motion from the handle to the brush. It required identifying whether the handles were manufactured without defects or if big dents, small dents, or some stripes could be appreciated. \textit{Iber-Oleff - Componentes Tecnicos Em Pl\'astico, S.A.} provided us with another dataset about automobile air vents they manufacture. The air vents have three components of interest: housing, lamellas (used to direct the air), and plastic links (which keep the lamellas tied together). The visual inspection task required us to determine whether (a) the fork is leaning against the support and correctly positioned, (b) the plastic link is present, (c) the lamella 1 is present, and the link is correctly assembled, and (d) the lamella 3 is present, and the link is correctly assembled.

Through the research, the researchers aimed to develop a comprehensive AI-first and human-centric approach to automated visual inspection.
In particular, they (i) developed machine learning models to detect defects, (ii) used active learning to enhance the models' learning process while alleviating the need to label data, (iii) used XAI to enhance the labeling process, (iv) analyzed how data augmentation techniques at embeddings and image level, along with anomaly maps can enhance the machine learning discriminative capabilities, (v) how human fatigue can be detected and predicted in humans, and (vi) how to calibrate and measure models' calibration quality to provide probabilistic predictive scores.

Research at the EU H2020 STAR project confirmed that active learning could alleviate the need for data labeling and help machine learning models learn better based on fewer data instances \cite{rovzanec2023active}. Nevertheless, the effort saved depends on the pool of unlabeled images, the use case, and the active learning strategy. Data augmentation techniques at an image or embedding level have increased the models' discriminative performance \cite{rovzanec2022synthetic}. Furthermore, complementing images with anomaly maps as input to supervised classification models has substantially improved discriminative capabilities \cite{rovzanec2022robust}. The data labeling experiments showed decreased labeling accuracy by humans over time \cite{rovzanec2023predicting}, which was attributed to human fatigue. While the future labeling quality can be predicted, it requires ground truth data. This can be acquired by showing synthetically generated images. Nevertheless, more research is required to devise new models that would consider other cues and predict human fatigue in data labeling without the requirement of annotated data. Finally, predictive scores alone provide little information to the decision-maker: predictive score distributions differ across different models. Therefore, performing probability calibration is paramount to ensure probability scores have the same semantics across the models. The research compared some of the existing probability calibration techniques and developed metrics to measure and assess calibration quality regardless of ground truth availability \cite{rovzanec2023active}.

\subsection{Human Digital Twins in Quality Control}\label{SS:HUMAN-DIGITAL-TWIN}

In the context of STAR, significant advancements have been made in developing human-digital twins (HDTs). In particular, the project has developed an infrastructure (Clawdite Platform \cite{montini2022iiot}) that allows the effortless creation of replicas of human workers through instantiating their digital counterparts. These HDTs have diverse features, encompassing static characteristics, dynamic data, and behavioral and functional models \cite{montini2021meta}.

To ensure a comprehensive representation of the human worker, STAR's HDT incorporates two crucial data types. Firstly, it assimilates physiological data collected from wearable devices. Secondly, it utilizes quasi-static data, which encapsulates characteristic attributes of the human, offering a holistic perspective on their traits. Central to STAR's HDT is an AI model designed to detect mental stress and physical fatigue. By leveraging physiological and quasi-static data, this AI model effectively gauges the stress and fatigue levels experienced by the human worker. This breakthrough in automated quality control holds remarkable significance, manifesting in two distinct ways:

\begin{itemize}
\item  During user manual inspection, the HDT continuously monitors the quality control process, actively identifying instances where the worker may be under significant mental or physical stress. In such cases, the system promptly suggests the worker take a break, ensuring their well-being and preventing any potential decline in performance.
\item During the training of automatic quality assessment models, as the worker evaluates and labels pictures during the data set creation, the system periodically assigns a confidence score to each label provided by the user. This confidence score is computed based on evaluating the worker's mental and physical stress levels estimated through the HDT's AI model. By considering these stress levels as an integral part of the quality evaluation process, the HDT provides valuable insights into the worker's state of mind and physical condition, allowing one to consider these features during the training of AI models for quality assessment and control. 
\end{itemize}

The integration of the HDTs, supported by the Clawdite Platform, in STAR's operations signifies a significant step forward in human-AI collaboration. This innovative approach prioritizes human workers' well-being and empowers automated quality control systems, ensuring optimal productivity and efficiency in various industrial settings.

\subsection{Making AI Visual Inspection Robust Against Adversarial Attacks}\label{SS:CYBERSECURITY}

In the context of the STAR project, an AI architecture was created for evaluating adversarial tactics and defense algorithms intended to safeguard, secure, and make the environments of manufacturing AI systems more reliable. More specifically, it was focused on AI-based visual inspection and tackled multiple use cases provided by two industrial partners: \textit{Philips Consumer Lifestyle BV} (Drachten, The Netherlands) and \textit{Iber-Oleff - Componentes Tecnicos Em Pl\'astico, S.A.} (Portugal). Current production lines are often tailored for the mass production of one product or product series in the most efficient way. Given its many advantages, AI is being increasingly adopted for quality inspection. Such models are usually trained considering some convolutional neural network (CNN), which then classifies whether a product is defective through inference upon receiving images captured by the inspection cameras. Nevertheless, such models can be attacked through adversarial data, leading AI models to wrongly classify the products (e.g., not detecting defects). For instance, the adversary may exploit a vulnerability in the visual inspection camera and compromise the integrity of the captured data by manipulating the operational behavior of this business resource.

\begin{figure}
\begin{center}
\includegraphics[width=0.5\textwidth]{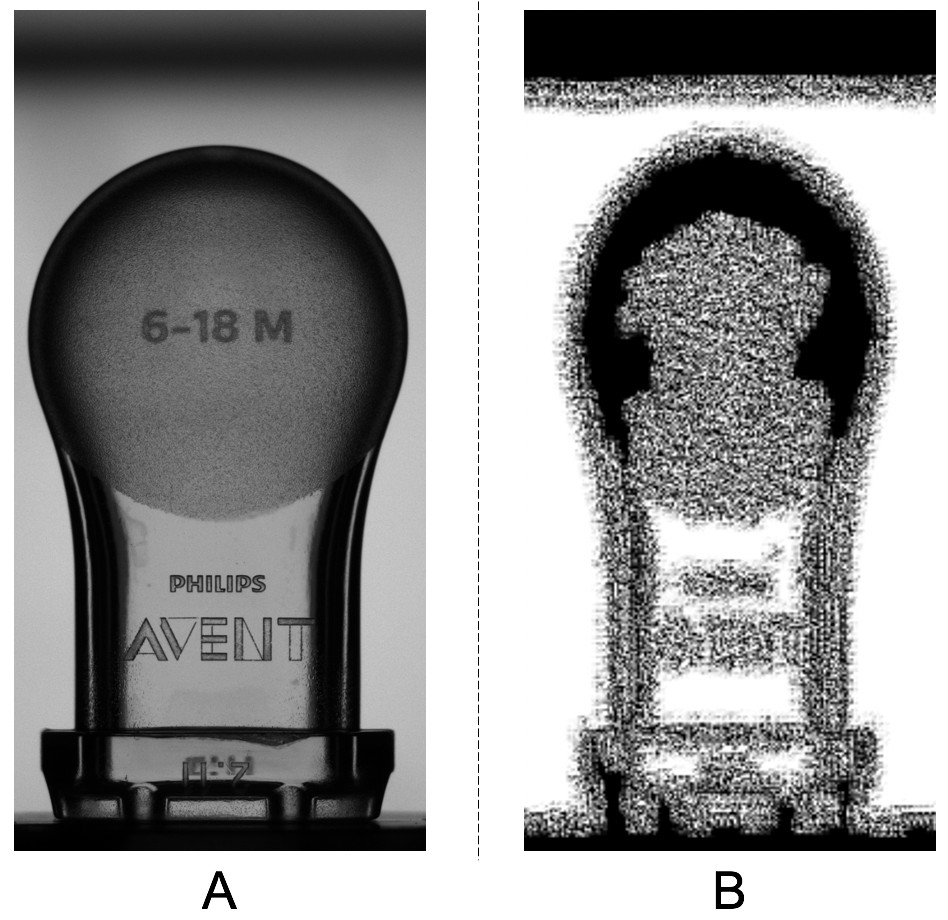}
\caption{Sample images. (A) corresponds to the original images received in the dataset regarding a soother cherry, while (B) corresponds to a perturbed image of a soother cherry. The DeepFool approach was used to generate perturbations in (B). The images are courtesy of Philips Consumer Lifestyle BV.} 
\label{F:CHERRIES}
\end{center}
\end{figure}

Among the various experimental testbeds built in the context of the STAR project, the ones created with soother cherries provided by \textit{Philips Consumer Lifestyle BV} (see Fig. \ref{F:CHERRIES}) were the most challenging. The cherry is the upper part of the soother. The high quality of the cherry must be guaranteed to avoid any harm to the babies. Therefore, detecting any adversarial attack is of primary importance, given the consequences of the attack can directly impact children's health. The goal of the testbed was to quantify the impact of adversarial attacks on classification models performing a visual inspection and evaluate how effective the defenses against such attacks were. To build the testbed, the Adversarial Robustness Toolbox \cite{nicolae2018adversarial} was used. In the experiments, the following adversarial methods were used: Fast Gradient Sign Attack (FGSM)  \cite{goodfellow2014explaining}, DeepFool  \cite{moosavi2016deepfool}, NewtonFool \cite{10.1145/3134600.3134635}, and Projected Gradient Descent (PGD) \cite{madry2017towards}. The aim was to utilize these well-documented adversarial methods to derive crafted instances that can be used to attack the baseline classification model. An example of a perturbed image using the Deepfool method is given in Fig.~\ref{F:CHERRIES}B. Experiments were performed with defence strategies, namely FeatureSqueezing \cite{xu2017feature}, JpegCompression \cite{dziugaite2016study}, SpatialSmoothing \cite{xu2017feature}, TotalVarMin \cite{guo2018countering}, and Adversarial Training  \cite{szegedy2013intriguing}.

\begin{figure}[h]
    \centering
    \includegraphics[width=\textwidth]{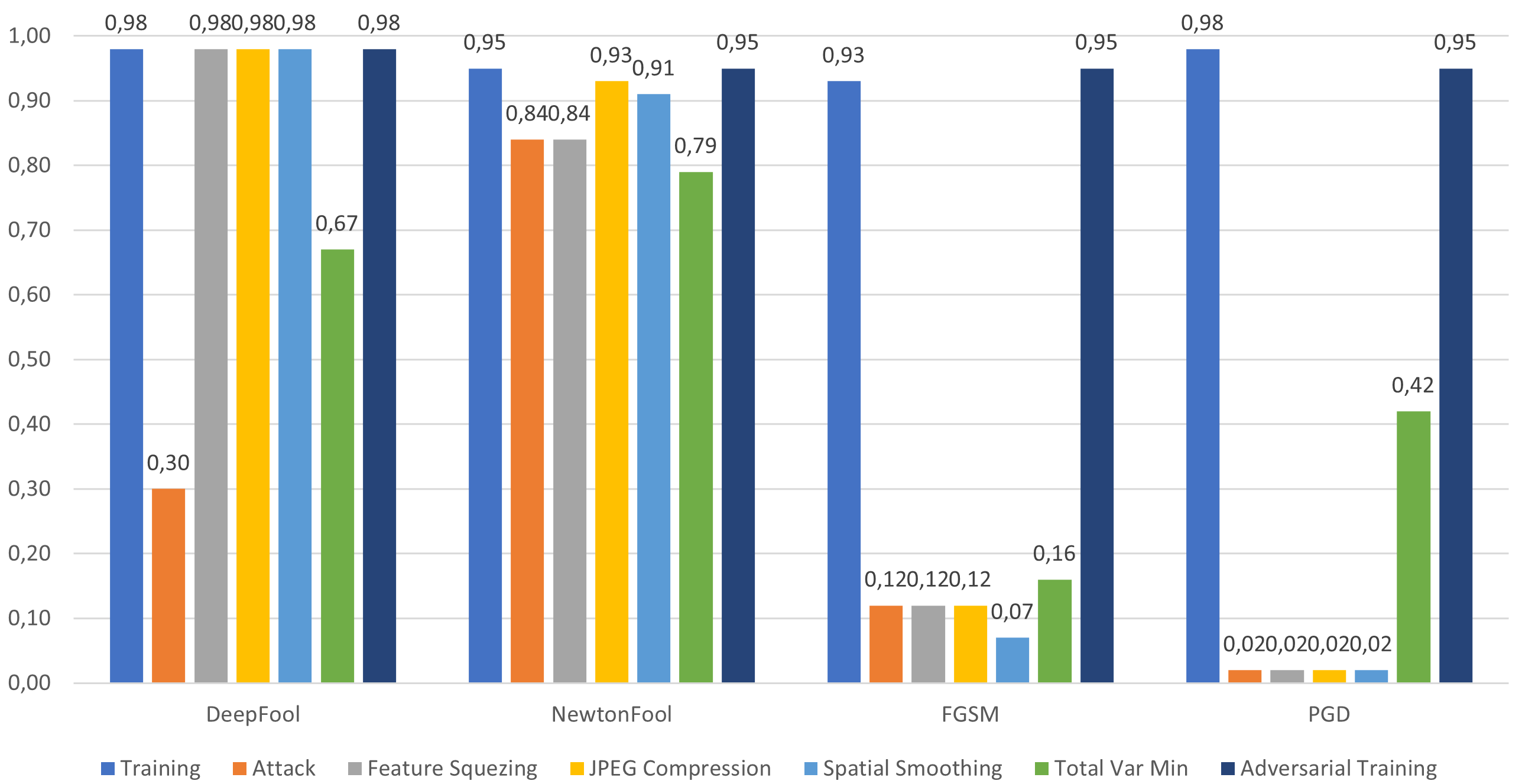}
    \caption{Evaluation results of pairwise comparison of adversarial attacks and defenses.}
    \label{F:RESULTS-ATTACKS}
\end{figure}

To gather insights regarding adversarial tactics and defenses, they were evaluated pairwisely. This enabled us to identify adversarial training as the best defense strategy to enhance the robustness of the CNN models. The basic idea behind adversarial training is to create examples that will be used later in the training process, creating a model aware of adversarial vectors launched against the quality control system. The results of the pairwise evaluation of the attacks and defenses are summarized in Fig. \ref{F:RESULTS-ATTACKS}. The results are grouped into four sets based on the attack strategy. A baseline classifier was initially trained for each of the four experiments (see tag "Training") to get the perception of the accuracy level that the quality inspection algorithm can achieve. The baseline model achieved an accuracy between 93\% and 98\%. The "Attack" bar indicates the accuracy of the classifier when posed against the adversarial attack. The DeepFool, FGSM, and PGD attacks strongly affected the classifier, causing the model's accuracy to drop below 30\%. This was not the case for the NewtonFool attack, where the classifier's accuracy dropped to 84\%. When considering defense strategies, Feature Squeezing, JPEG Compression, and Spatial Smoothing can defend against the DeepFool attack: for the given dataset, they led to an accuracy of 98\%. However, TotalVarMin failed to defend the model. All the defenses failed against the FGSM and the PGD attacks. Based on the acquired results of the pairwise evaluations, it became clear that no clear mapping exists between types of attacks and defenses. Therefore it can be challenging for defenders to plan a strategy to cope against any attack successfully. This outcome advocates the criticality and the challenge of defending against adversarial AI attacks. While the off-the-shelf and state-of-the-art defenses cannot perform in a stable manner under different adversarial methods, the Adversarial Training approach seems robust. The results agree with the literature, advocating that Adversarial Training can be a robust solution that can cope with adversaries despite its simplicity. A more detailed description of the abovementioned work can be found in \cite{s22186905}.

\section{Conclusion}\label{S:CONCLUSION}
This work has briefly introduced state-of-the-art research on human-machine collaboration, perspectives on human-centric manufacturing, and the key aspects of trustworthiness and accountability in the context of Industry 5.0. It described research on automated quality inspection, considering the role of robotics, AI approaches, and solutions to visual inspection and how a fruitful human-machine collaboration can be developed in the visual inspection domain. Finally, it described the experience and results obtained through research performed in the EU H2020 STAR project. 

The converging view from the literature analysis is that human-machine cooperation requires adequate communication and control realized through effective bidirectional information exchange. Studies have been performed to understand peoples' emotional and social responses in human-machine interactions, understand task design, and how humans' trust, acceptance, decision-making, and accountability are developed or impacted in the presence of machines. In the field of visual inspection, much research was invested in automating the task of visual inspection by developing machine learning models to detect product defects. Furthermore, many research efforts targeted the development of techniques for XAI related to machine vision. Visual aids and hints derived by XAI are conveyed to humans through heat maps. Similarly, insights obtained from unsupervised machine learning models are conveyed to humans as anomaly maps. While such approaches solve particular problems, little research describes how a human-in-the-loop approach could be developed for visual inspection in manufacturing settings. This research aims to bridge the gap by implementing existing and researching novel active learning techniques for data selection to enhance the learning of machine learning algorithms. It also explores how labeling requirements could be reduced by employing few-shot learning and active learning techniques. Furthermore, research was conducted to understand how XAI and unsupervised classification methods can be used to generate heatmaps and anomaly maps to facilitate data labeling in the context of manual revision or data annotation tasks. Moreover, predictive models were developed to predict how heat maps and anomaly maps should be adapted over time to bridge the gap between the information conveyed by machine learning algorithms and explainability techniques and human perception. In addition, experiments were performed to gain insights related to human-fatigue monitoring in the context of visual inspection. The present work described a complete and modular infrastructure developed to instantiate HDT, and different AI models for perceived fatigue exertion and mental stress have been trained to derive relevant features for human-centered production systems. Finally, it describes some research on adversarial attacks and defenses to enhance the understanding of protecting visual inspection setups in manufacturing environments.

While the research presented above advances the understanding of developing a human-in-the-loop approach for visual inspection in manufacturing, many open issues remain to be solved. Further research is required to understand how adaptive humans perceive hinting and how the many solutions described above contribute to building trust between humans and machines. Furthermore, effort must be invested to quantify the benefits such solutions bring to a manufacturing plant when implemented. Future research will encompass the integration of these solutions, aiming to achieve a comprehensive and synergistic implementation. The research will aim to develop new approaches that interleave active learning and XAI. Furthermore, novel few-shot learning solutions will be considered to allow for greater flexibility of the visual inspection while reducing data labeling requirements to a minimum. Finally, integrating AI visual inspection models and the HDT is expected to significantly augment the efficacy of quality inspection processes during user manual assessment and AI model training.

\bmhead{Acknowledgments}

This work was supported by the Slovenian Research Agency and the European Union's Horizon 2020 program project STAR under grant agreement number H2020-956573.


\bibliography{main}% common bib file

\end{document}